\documentclass[10pt,conference]{IEEEtran}
\usepackage{dcolumn}
\usepackage{fancyheadings}
\usepackage{amstext,amssymb}
\usepackage{graphicx}
\usepackage{times,color}
\usepackage[final,ulem=normalem]{changes}
\usepackage{subfig}
\usepackage{algorithm}
\usepackage{algorithmic}
\usepackage{latexsym}
\usepackage{amsmath}
\usepackage{flushend}
\newtheorem{theorem}{Theorem}
\newtheorem{lemma}{Lemma}
\newtheorem{corollary}{Corollary}

\newtheorem{definition}{Definition}
\newtheorem{remark}{Remark}

\newcommand{\comment}[1]{}

\definecolor{dgreen}{rgb}{0,0.48,0.3}

\long\def\comment#1{}

\IEEEoverridecommandlockouts

\definechangesauthor{RW}{blue}
\definechangesauthor{JW}{red}
\definechangesauthor{LD}{dgreen}
\definechangesauthor{ZZ}{orange}
\begin{document}
\pagestyle{empty}
\title{Absorbing Set Spectrum Approach for Practical Code Design}
\author{Jiadong Wang, Lara Dolecek, Zhengya Zhang and Richard Wesel\\
wjd@ee.ucla.edu,~dolecek@ee.ucla.edu,~zhengya@eecs.umich.edu,~wesel@ee.ucla.edu
\thanks{This research was supported by a gift from Inphi Corp. and in part by grant CCF-1029030 from NSF. }}

\maketitle \thispagestyle{empty}
\begin{abstract}
\deleted[RW]{\added[JW]{PLEASE HELP WITH THIS PART.} Regular LDPC codes tend to have better error-floor behavior than irregular LDPC codes.  However, for moderate block lengths and high rates, the error floor remains a concern even for regular LDPC codes.  This is especially the case for applications such as \deleted[ZZ]{memory media}\added[ZZ]{data storage} that require very low \deleted[ZZ]{frame} error rates\deleted[ZZ]{(FERs)}.  }This paper focuses \added[ZZ]{on} controlling \added[ZZ]{the} absorbing set\deleted[ZZ]{s} \added[ZZ]{spectrum} for a class of regular LDPC codes\deleted[ZZ]{:}\added[ZZ]{ known as} separable, circulant-based  (SCB) codes. For a specified circulant matrix, SCB codes all share a common mother matrix\deleted[ZZ]{and include}\added[ZZ]{, examples of which are} array-based LDPC codes and many common quasi-cyclic codes.  SCB codes retain \added[ZZ]{the} standard properties of quasi-cyclic LDPC codes such as girth, code structure, and compatibility with \deleted[ZZ]{existing high-throughput hardware}\added[ZZ]{efficient decoder} implementations. \added[ZZ]{In} this paper\added[ZZ]{, we define} \deleted[ZZ]{uses} a cycle consistency matrix (CCM) for each absorbing set of interest in an SCB LDPC code.  For an absorbing set to be present in an SCB LDPC code, the associated CCM must {not} be full {column-}rank. Our approach selects rows and columns from the SCB mother matrix to systematically eliminate dominant absorbing sets by forcing the associated CCMs to be full {column-}rank.  \deleted[ZZ]{This paper uses}\added[ZZ]{We use} the CCM approach to select rows from the SCB mother matrix to design SCB codes \deleted[ZZ]{for $r$$=$$5$}\added[ZZ]{of column weight $5$} that avoid all \added[ZZ]{low-weight absorbing sets} $(4,8)$, $(5,9)$, and $(6,8)$\deleted[ZZ]{ absorbing sets}.  Simulation results demonstrate that the newly designed code has a steeper error-floor slope and provides at least one order of magnitude of improvement in the low \deleted[ZZ]{FER}\added[ZZ]{error rate} region as compared to an elementary array-based code. \deleted[RW]{Identifying absorbing-set-spectrum equivalence classes within the family of SCB codes with a specified circulant matrix significantly reduces the search space of possible code matrices.}
\end{abstract}


\section{Introduction}


\added[LD]{It is well known that finite-length low-density parity-check (LDPC) codes suffer performance degradation  in the high signal-to-noise \added[ZZ]{ratio} (SNR)/low frame error rate (FER) region. This degradation is commonly referred to as the error floor. Prior work indicate\deleted[ZZ]{s}\added[ZZ]{d} that certain sub-graphs called trapping sets~\cite{richardson}, and, in particular, a subset of trapping sets called absorbing sets~\cite{dolecekIT10} are a primary cause of the error floor. \replaced[RW]{A}{Combinatorially defined, a}bsorbing set is a particular type of a trapping set\deleted[RW]{s} that is  stable under bit-flipping decoding. \replaced{This paper improves performance by controlling absorbing sets for}{In this paper we focus on the performance of} a class of regular LDPC codes\added[RW]{, known as separable circulant-based (SCB) codes,} that \replaced{are constructed as an arrangement}{are built out} of circulant matrices.\deleted[RW]{,  cast in terms of their absorbing set properties}}

Recent papers have proposed methods to lower the error floor by improving the absorbing set (or trapping set) spectrum.
For example, small trapping sets can be avoided by introducing additional check nodes~\cite{milenkovicGlobecom06}, or \added[LD]{by} increasing the girth{~\cite{MilenkovicIT06}}.   The algorithm in~\cite{NguyenITW10} constructs quasi-cyclic codes from Latin squares so that the Tanner graph \added[LD]{of the code} does not contain certain trapping sets. \deleted[LD]{\added[JW]{Recent results in \cite{SchlegelZhangIT10}, \cite{KyungISIT10}, \cite{ZhengTCOM10} also explore the error floor with the trapping set spectrum.}}\added[LD]{Recent results \cite{SchlegelZhangIT10}, \cite{KyungISIT10}, \cite{ZhengTCOM10} have also investigated the errors floor of certain practical codes in terms of their trapping/absorbing sets.}

\added[LD]{A recently proposed approach \cite{DOLECEKISTC10, JWANGICC11} avoids certain dominant absorbing sets without compromising code properties by carefully selecting the rows/columns of the (SCB) mother matrix.  This paper builds upon the cycle consistency matrix (CCM) approach \cite{WangITA2011} to analyze \deleted[RW]{a} SCB codes with column weight 5,} \added[RW]{i.e., codes with five rows of circulant submatrices.}

\added[LD]{For an absorbing set to be present in an SCB LDPC code, the associated CCM must not be full column-rank.  \replaced[RW]{Furthermore, analysis of}{By exploring} the variable-node graph \replaced{for a  variety of}{among different} absorbing sets\replaced[RW]{ reveals that}{,} the existence of certain absorbing sets is \replaced{a}{the} necessary condition \replaced{for}{of} other \added[RW]{(larger)} absorbing sets \added[LD]{to exist}. Using these \added[RW]{two} observations, \replaced[RW]{this paper analyzes}{we analyze} the smallest absorbing sets in \replaced[RW]{$r$$=$$5$}{this class of} SCB codes, and systematically avoids these absorbing sets by selecting rows from the SCB mother matrix \replaced{that force}{and forcing} the associated CCMs to be full column-rank. \added[RW]{FPGA} simulation results \deleted[ZZ]{on} \deleted[RW]{FPGA }\deleted[ZZ]{support our analysis and} \added[ZZ]{confirm}\deleted[ZZ]{confirm} that the new codes have significantly steeper error-floor slopes \deleted[JW]{in the low-FER region}.}

\added[LD]{As a representative instance of SCB codes, Tanner-construction codes \cite{tanner04} with moderate rates are analyzed. It is shown that these codes can have \deleted[ZZ]{good}\added[ZZ]{improved} absorbing set spectr\deleted[ZZ]{um}\added[ZZ]{a} by carefully avoiding the smallest absorbing sets with suitable code parameter selection. 
}

Section~\ref{background} describes separable circulant-based (SCB) codes and the cycle consistency matrix (CCM). Section~\ref{analysis5} applies the CCM approach to analyze three groups of dominant absorbing sets in an example family of SCB codes. Section~\ref{analysis5} then selects specific rows from the SCB mother matrix to eliminate all the dominant absorbing sets by forcing the associated CCMs to be full column-rank. Section~\ref{analysis5} also \added[LD]{analyzes} the absorbing set\deleted[RW]{s} spectrum of Tanner-construction codes and provides several good row-selection functions. Section~\ref{results} provides \added[ZZ]{hardware} simulation results demonstrating the performance improvement \deleted[ZZ]{\added[JW]{in hardware} obtained} \replaced[RW]{obtained with}{by} the new codes. Section~\ref{conclusion} delivers the conclusions.

\vspace{-0.0in}\section{Definition and Preliminaries}\label{background}
This section introduces separable, circulant-based (SCB) codes and the cycle consistency matrix (CCM) associated with absorbing sets in SCB codes.
\subsection{Circulant-based LDPC codes}
Circulant-based LDPC codes are a family of structured regular $(r,c)$ codes where $r$ is the variable-node degree and $c$ is the check-node degree.  They are constructed as $r$ rows and $c$ columns of circulant matrices. They are \added[ZZ]{especially amenable to}\deleted[ZZ]{particularly compatible with} high-throughput hardware implementations~\cite{zhangTCOM}.

The parity-check matrix of circulant-based LDPC codes has the following general structure:
\small
\begin{equation*}
H_{p,f}^{r,c}=\left[\begin{array}{ccccc}
\sigma^{f(0,0)} & \sigma^{f(0,1)} & \sigma^{f(0,2)} & \ldots & \sigma^{f(0,c-1)}\\
\sigma^{f(1,0)} & \sigma^{f(1,1)} & \sigma^{f(1,2)} & \ldots & \sigma^{f(1,c-1)}\\
\sigma^{f(2,0)} & \sigma^{f(2,1)} & \sigma^{f(2,2)} & \ldots & \sigma^{f(2,c-1)}\\
\vdots & \vdots & \vdots & \ldots & \vdots \\
\sigma^{f(r-1,0)} & \sigma^{f(r-1,1)} & \sigma^{f(r-1,2)} & \ldots & \sigma^{f(r-1,c-1)}\\
\end{array}
\right]~,
\end{equation*}\normalsize
where $ \sigma $ is a $ p \times p $ circulant matrix.

A column (row) group is a column (row) of circulant matrices. Each variable node has a label $(j,k)$ with $j\in \{0,..., c-1\}$ \replaced[RW]{being}{that is} the index of the corresponding column group and $k\in \{0,..., p-1\}$ \replaced[RW]{identifying}{identifies} the specific column within the group. Similarly, each check node has a label $(i,l)$ where $i\in \{0,..., r-1\}$ and $l\in \{0,..., p-1\}$.

Circulant-based LDPC codes include, for example, the constructions in{~\cite{tanner04,FossorierIT04} and~\cite{lincostello17}}. The girth can be guaranteed to be at least 6 by placing a constraint on the values of {the submatrix exponent} $f(i,j)$~\cite{dolecekIT10}.

This paper focuses on separable, circulant-based (SCB) codes defined as follows:

\begin{definition}[Separable, Circulant-Based (SCB) Code]
An {SCB} code is a circulant-based LDPC code with a parity-check matrix $H_{p,f}^{r,c}$ in which  $ f(i,j) $ is separable, i.e., $f(i,j)=a(i) \cdot b(j)$.\hfill$\blacksquare$
\end{definition}

Parity check matrices of SCB codes with the specified circulant matrix can be viewed as originating from  a common SCB mother matrix $H_{p,f_m}^{p,p}$ with $ f_m(i,j)={i \cdot j} $.  The functions $a(i)$ and $b(j)$ effectively specify which rows and columns of the mother matrix are selected for the resultant SCB matrix.  The ranges of $a(i)$ and $b(j)$ are $\{0, \ldots, p-1\}$.

The SCB structure imposes \replaced[RW]{four}{4} conditions \cite{dolecekIT10} on the variable and check nodes: (1) bit consistency, (2) check consistency, (3) pattern consistency, (4) cycle consistency. These conditions are essential to the CCM approach that is introduced next.

%
%
%
%


\subsection{Absorbing sets and the Cycle Consistency Matrix} \label{mtrxrep}
An LDPC code with parity-check matrix $ H $ is often viewed as {a {bipartite ({Tanner}) graph} $G_H = (V, F,E)$}, where the set $V$ represents the variable nodes, the set $F$ represents the check nodes, and $E$ corresponds to the edges between variable and check nodes.

\added[LD]{For a variable node subset $V_{\text{as}} \subset V$, let $G_{\text{as}} = (V_{\text{as}}, F_{\text{as}},E_{\text{as}})$ be the bipartite graph of the edges $E_{\text{as}}$ between the variable nodes $V_{\text{as}}$ and their neighboring check nodes $F_{\text{as}}$.  Let $o(V_{\text{as}})\subset F_{\text{as}}$ be the neighbors of $V_{\text{as}}$ with odd degree (unsatisfied check nodes) in $G_{\text{as}}$  and $e(V_{\text{as}})\subset F_{\text{as}}$ be the neighbors of $V_{\text{as}}$ with even degree in $G_{\text{as}}$ (satisfied check nodes).}
\added[LD]{An $(a,b)$ \textit{absorbing set}~\cite{dolecekIT10} $G_{\text{as}} = (V_{\text{as}}, F_{\text{as}},E_{\text{as}})$ is a Tanner graph with $a$ variable nodes\replaced{,}{ and} $b$ odd-degree check nodes, and \added[RW]{with} each variable node \replaced[RW]{having}{has} strictly \replaced[RW]{fewer}{less} odd-degree neighbors than even-degree neighbors.}

%
%

Suppose there are $n$ variable nodes in the absorbing set. Let $j_1, \ldots, j_n$ be the column-group labels of these $n$ nodes in the SCB mother matrix. Define $u_m=j_{m}-j_1 ,m=2,...,n$ and $ \mathbf{u}=[u_2,...,u_{n}] $. For each cycle in the absorbing set, by replacing the difference of $ j $'s with the difference of $ u $'s \deleted[LD]{and manipulating the expression}, we can rewrite the cycle consistency equation as
\small
\vspace{-0.1in}\begin{equation} \label{equcycleu}
\sum\limits_{m = 2}^{t} (i_{m-1}-i_{m})u_{m}   = 0 \mod p  ,
\end{equation}
\normalsize
where $2t$ is the cycle length.  Note that $i_m$ will be different for different cycles reflecting the particular cycle trajectories.

Every cycle in the absorbing set satisfies an equation of the form \eqref{equcycleu}.  Taken together, these equations produce a matrix equation: $ \mathbf{M}\mathbf{u} = 0 \mod p $, where $ \mathbf{M}_{ym} $ is the coefficient of $ u_m $ in \eqref{equcycleu} for the $ y $th cycle.

A key property of $\mathbf{M}$ is that $ \mathbf{M}\mathbf{u} = 0 \mod p $ completely characterizes the requirement that every cycle in $G_{\text{as}}$ satisfies \eqref{equcycleu}.  Even so, it is not necessary for $\mathbf{M}$ to include a row for every cycle in the absorbing set.

A cycle need not be included in $\mathbf{M}$ if it is a linear combination of cycles already included in $\mathbf{M}$. Thus the number of rows needed in $\mathbf{M}$ is the number of linearly independent cycles in $G_{\text{as}}$.  \replaced[RW]{Two}{Some} definitions~\cite{Diestelgraphtheory} from graph theory are necessary to establish the number of linearly independent cycles in $G_{\text{as}}$ and hence how many rows are needed for $\mathbf{M}$.

\begin{definition}[Incidence Matrix]
 For a graph with $n$ vertices and $q$ edges, the (unoriented) incidence matrix  is an $n \times q$  matrix $B$ with $ B_{ij}=1 $ if vertex $ v_i $ and edge $ x_j $ are incident and 0 otherwise.\hfill$\blacksquare$
\end{definition}

%
\begin{definition}[Binary Cycle Space]
The binary cycle space of a graph is the null space of its incidence matrix over $ GF(2) $.\hfill$\blacksquare$
\end{definition}

Any absorbing-set bipartite graph $G_{\text{as}}$ can be transformed into a graph whose only vertices are $V_{\text{as}}$ and where two vertices are connected iff there is a check node that connects them. We call this graph the \textit{variable-node (VN) graph} of the absorbing set.
The incidence matrix \added[RW] of the VN graph provides a characterization of all the cycles in an absorbing set.
The number of linearly independent cycles in an absorbing set, which is the dimension of its binary cycle space, is the size of the null space of the incidence matrix $B_\text{as}$: $D_{\text{bcs}}=q-\text{rank}(B_\text{as})$.

Having established the number of rows in $\mathbf{M}$, it can be formally defined as the Cycle Consistency Matrix:
\begin{definition}[Cycle Consistency Matrix]
The cycle consistency matrix $\mathbf{M}$ of an absorbing-set graph $G_\text{as}$ has $|V_{\text{as}}|-1$ columns and $D_{\text{bcs}}$  rows. The rows of $\mathbf{M}$ correspond to $D_{\text{bcs}}$ linearly independent cycles in $G_\text{as}$.
Each row has the coefficients of $\mathbf{u}$ in \eqref{equcycleu} for each \deleted[RW]{one }of these linearly independent cycles. \hfill$\blacksquare$\end{definition}

Note that  $\mathbf{M} \cdot \mathbf{u} = 0 \mod p$ completely characterizes the requirement that every cycle in $G_{\text{as}}$ satisfies \eqref{equcycleu}.

The vector $\mathbf{u}$ cannot be an all-zero vector because an all-zero $\mathbf{u}$ indicates that all variable nodes have the same column group.  This violates the Check Consistency condition, which requires that variable nodes sharing a check node have distinct column groups.  Thus $\mathbf{u} \neq \mathbf{0}$, and a necessary condition for the existence of a given absorbing set is that its $\mathbf{M}$ does not have full {column-}rank in $GF(p)$.

If the VN graph of the absorbing set $G_\text{as}^A$ is a sub-graph of the VN graph of another absorbing set $G_\text{as}^B$ with the same number of variable nodes, then we say the VN graph of the absorbing set $G_\text{as}^A$ is extensible.

\begin{theorem} \label{thm:ccm}
Given a proposed absorbing set graph $G_{\text{as}} = (V_{\text{as}}, F_{\text{as}},E_{\text{as}})$, {where every variable node is involved in at least one cycle}\footnote{{\deleted[LD]{It is easy to show if} \added[LD]{If }the variable node degree is at least 2, then each variable node in a given absorbing set must be a part of at least one cycle.}}, \replaced[RW]{specified}{the} column group labels of the variable nodes in $V_{\text{as}}$ in the SCB mother matrix, and \replaced[RW]{specified}{the} row-group labels of the check nodes in $F_{\text{as}}$ in the SCB mother matrix, the following are necessary conditions for the proposed absorbing set to exist in each daughter SCB LDPC code (with a parity check matrix $H$ that includes the specified row and column groups of that SCB mother matrix):
(1) The CCM for $G_{\text{as}}$ {does not have full column-rank};
(2) Variable nodes in  $V_{\text{as}}$ satisfy the Bit Consistency condition and can form a difference vector ${\bf u}$ in the null space of the CCM; and
(3) Each check node in  $F_{\text{as}}$ satisfies the Check Consistency condition.
Taken together, these conditions are also sufficient if the VN graph of this absorbing set is not extensible.
\end{theorem}

\textit{Proof:}
\replaced[RW]{The}{For the lack of space} \added[LD]{proof is provided in Appendix~\ref{pr:thmccm}.}
\hfill$\blacksquare$

\begin{corollary} \label{cor:ext}
Given an $ (a_1,b_1) $ absorbing set graph $G^{1}_{\text{as}} = (V^{1}_{\text{as}}, F^{1}_{\text{as}},E^{1}_{\text{as}})$ and an $ (a_2,b_2) $ absorbing set graph $G^{2}_{\text{as}} = (V^{2}_{\text{as}}, F^{2}_{\text{as}},E^{2}_{\text{as}})$, if $ a_1 \leq a_2 $ and the VN graph of $G^{1}_{\text{as}}$  is a sub-graph of the VN graph of $G^{2}_{\text{as}}$, then the existence of $G^{1}_{\text{as}}$ is a necessary condition of the existence of $G^{2}_{\text{as}}$.
\end{corollary}

\textit{Proof:}
\deleted[RW]{Proof is provided in }\added[LD]{Appendix~\ref{pr:corext}} \added[RW]{provides the proof.}
\hfill$\blacksquare$

\section{Illustrative case study with $ r=5 $}\label{analysis5}
The approach in \cite{DOLECEKISTC10} reveals that a careful selection of $ r $ row-groups from the SCB mother matrix can eliminate certain small absorbing sets to improve the error floor. This section provides an example with $r=5$ (five row groups) that shows how to use the new CCM approach to efficiently improve an SCB code by analytically avoiding \deleted[ZZ]{more}\added[ZZ]{low-weight} absorbing sets with careful row selections.

Our example of SCB code design involves three classes of SCB codes: (1) Array-based codes~\cite{fan}: the most elementary SCB codes in which the first $r$ rows of the SCB mother matrix $H_{p,f}^{p,p}, f(i,j)=i \cdot j $ comprise the parity-check matrix. We will refer to this class as the elementary array-based (EAB) codes. (2) Selected-row (SR) SCB codes: the parity-check matrix for these codes is $H_{p,f}^{r,p}, f(i,j)=a(i) \cdot j $ where $a(i)$ is called the row-selection function (RSF). (3) Shortened SR (SSR) SCB codes: the parity-check matrix for these codes is $H_{p,f}^{r,c}, f(i,j)=a(i) \cdot b(j) $ where $b(j)$ is called the column-selection function (CSF).

Theorem~\ref{thm:ccm} shows that a\added[RW]{n} absorbing set \replaced[RW]{may}{can} be avoided \added[RW]{either} by forcing the associated CCM to be full column rank or \added[RW]{by} preclud\replaced{ing}{e} $\mathbf{u}$ from \added[RW]{being in} the null space of $\mathbf M$. Corollary~\ref{cor:ext} shows that if a\added[RW]{n} \deleted[RW]{given }absorbing set does not exist, then \replaced[RW]{all}{any} absorbing set\added[RW]{s} whose VN graph\added[RW]{s} contain\deleted[RW]{s} the VN graph of this absorbing set \added[RW]{also} do\deleted[RW]{es} not exist. \replaced[RW]{The}{Thus our} CCM approach carefully selects \added[RW]{the} RSF and CSF to systematically eliminate small absorbing sets, in the order of the size of the VN graph of the absorbing sets.


\replaced[RW]{Prior}{A previous} result\added[RW]{s}~\cite{dolecekITA10} prove\deleted[RW]{s that} $(4,8)$ absorbing sets \replaced[RW]{to be}{are} the smallest possible for a general $r$$=$$5$ SCB code family\replaced{ and to dominate}{, and this class of absorbing sets dominate} the low BER region~\cite{zhangTCOM}. \replaced[RW]{H}{The h}ardware simulation results in \cite{DOLECEKISTC10} show\deleted[RW]{s} that the next two dominant absorbing sets in $r$$=$$5$ EAB codes or SR SCB codes are $ (5,9) $ and $ (6,8) $.

\replaced[RW]{Careful selection of the RSF via the CCM approach successfully avoids all $ (4,8) $, $ (5,9), $ and $ (6,8) $ absorbing sets.}{We will use CCM approach by carefully choosing RSF to avoid all $ (4,8) $, $ (5,9) $ and $ (6,8) $ absorbing sets successively.}
Sections~\ref{subsection48}, \replaced[RW]{~\ref{subsection59}, and \ref{subsection68}}{Section~\ref{subsection59} and Section~\ref{subsection68}}, respectively, show that $(4,8)$, $(5,9)$, and $(6,8)$ absorbing sets exist in \added[RW]{the} EAB SCB code with $r$$=$$5$ \replaced{but also that SR SCB codes can systematically eliminate these configurations with}{and} \replaced{a carefully selected RSF} {carefully selecting row groups from the SCB mother matrix can systematically eliminate}. Section~\ref{nonexistence3sets} provides \replaced[RW]{example RSFs}{several good row-mapping vector examples} that\deleted[RW]{can} eliminate small absorbing sets with $r$$=$$5$. Section~\ref{tannerfor5} explores the absorbing set spectrum of the existing quasi-cyclic LDPC codes with \added{the} Tanner\deleted[RW]{'s} construction~\cite{tanner04}.

\subsection{$ (4,8) $ absorbing sets} \label{subsection48}
\begin{figure}[!t]
\centerline{\subfloat[$(4,8)$ absorbing set]{\includegraphics[width=0.2\textwidth]{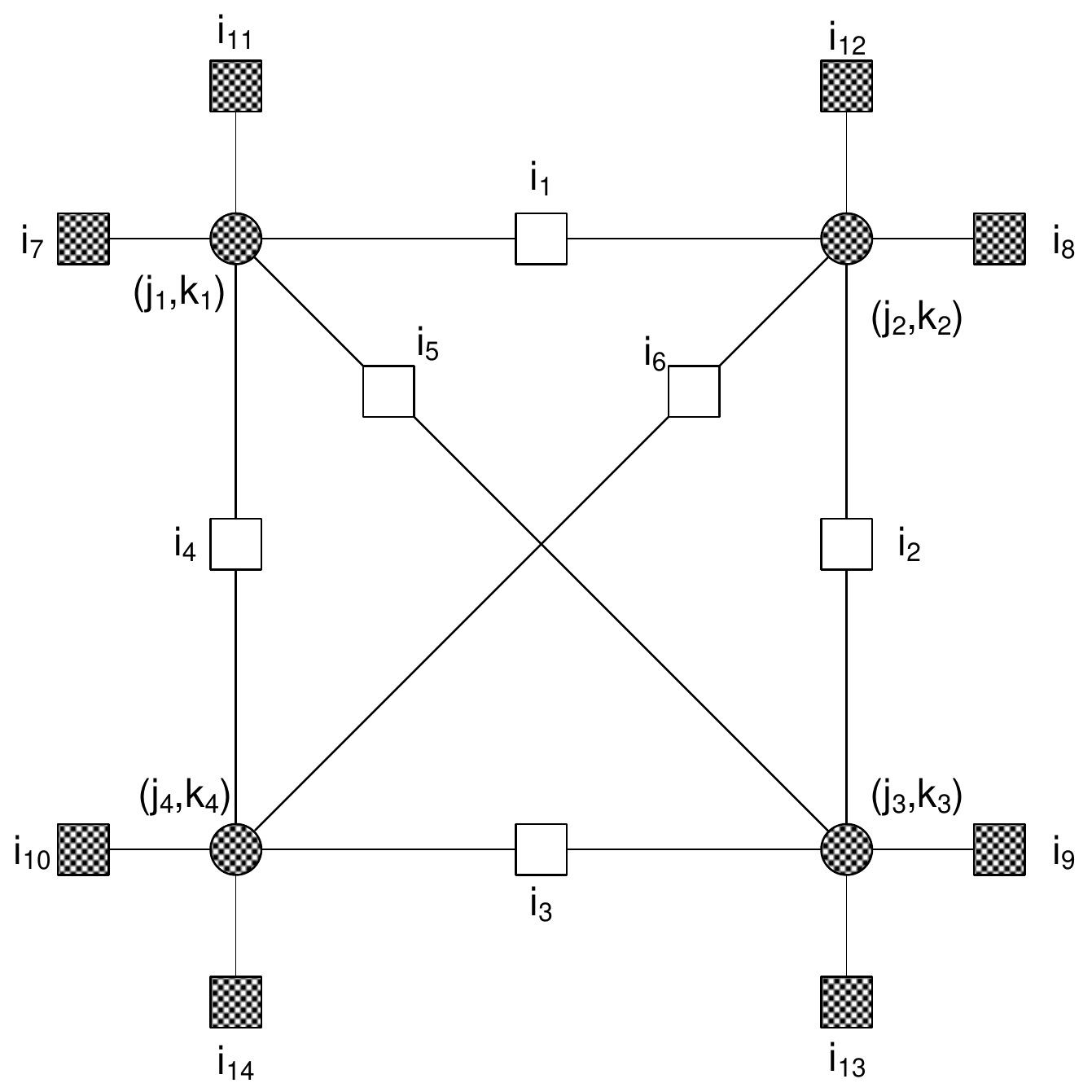}
\label{fig48}}
\hfil
\subfloat[$(5,9)$ absorbing set]{\includegraphics[width=0.2\textwidth]{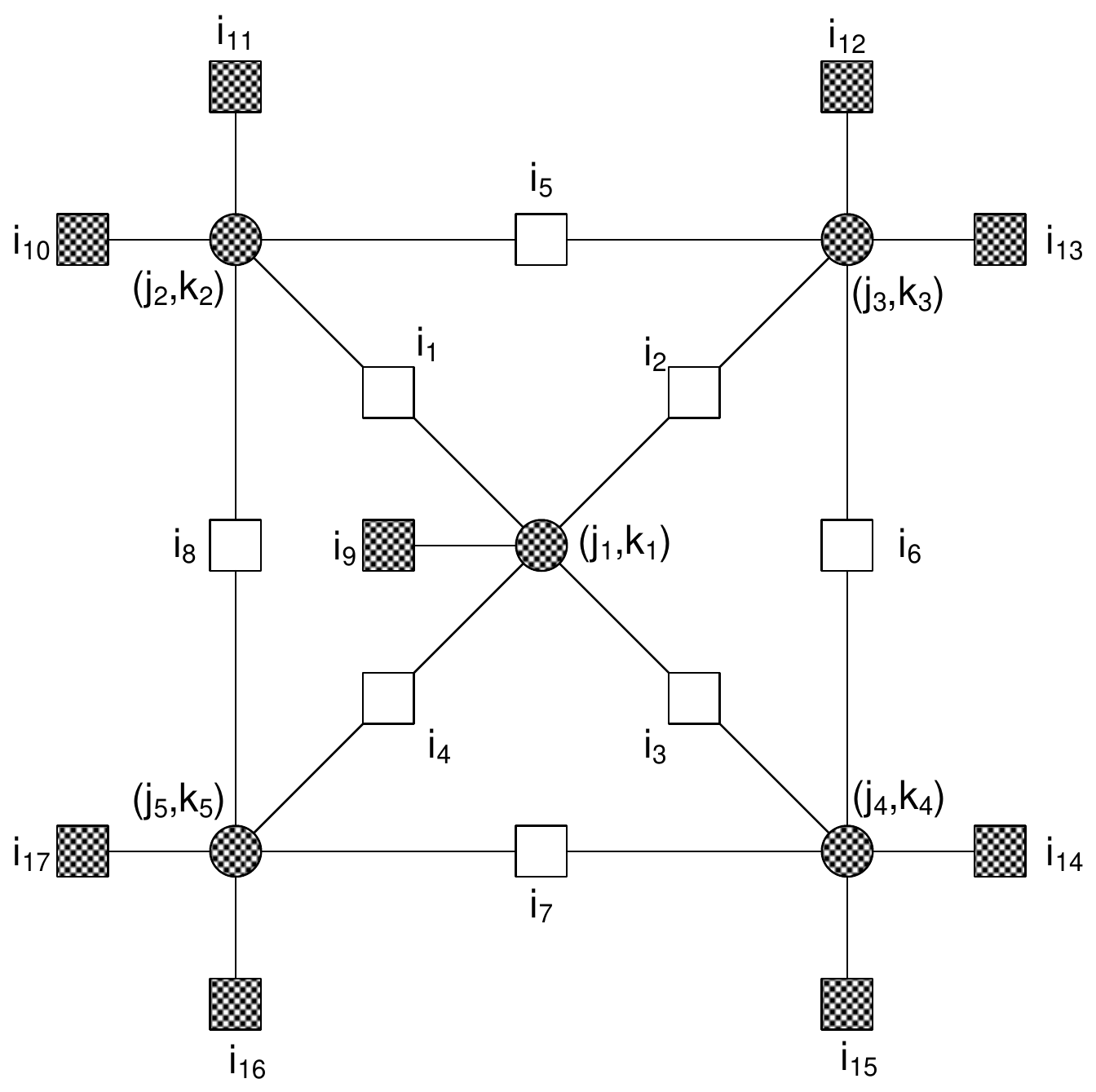}
\label{fig59}}}
\caption{Depiction of absorbing sets.}\vspace{-0.3in}
\end{figure}

\added[LD]{A result in~\cite{dolecekITA10} shows that $(4,8)$ absorbing set in \added[LD]{Fig.~\ref{fig48}} is the smallest absorbing set in the $H_{p,f(i,j)}^{5,p}$ code family, and~\cite{DOLECEKISTC10} shows that certain RSF\added[RW]{s} \deleted[RW]{can} eliminate the $(4,8)$ absorbing sets in SR SCB codes. This section \replaced{uses the}{applies} CCM approach to \replaced[RW]{efficiently establish}{systematically build the} necessary and sufficient conditions for the existence of $ (4,8) $ absorbing sets.}
\begin{lemma}\label{lemma48}
$\det \mathbf{M}=0 \mod p$ is a necessary and sufficient condition for the existence of $ (4,8) $ absorbing sets in Fig.~\ref{fig48}.
\end{lemma}

\textit{Proof}:
\added[LD]{Appendix~\ref{pr:lemma48} provides the proof.}
\hfill$\blacksquare$

\replaced[RW]{Corollaries \ref{cor:eab48} and \ref{cor:sr48} restate Lemmas 1 and 2 of \cite{DOLECEKISTC10}.  The CCM approach concisely proves these corollaries in Appendix \ref{pr:cor2and3}.}{Following the analysis of Lemma 1 and 2 in \cite{DOLECEKISTC10}, we can reach the analogous conclusion with following two corollaries.}
\begin{corollary} \label{cor:eab48}
$(4, 8)$ \deleted[RW]{(fully)} absorbing sets exist in EAB codes described by the parity check matrix $H_{p,i \cdot j}^{5,p}$, and their number scales as $\Theta(p^3)$.
\end{corollary}
\begin{corollary} \label{cor:sr48}
There are no $(4,8)$ absorbing sets in the SR SCB codes described by the parity check matrix $H_{p, a(i)\cdot j}^{5,p}$, for prime $p$ large enough with \added[LD]{a} proper choice of RSF.
\end{corollary}

\begin{remark}
The SR SCB codes avoid $ (4,8) $ absorbing sets by carefully choosing RSF such that $ \det \mathbf{M} \neq 0 \mod p $ for $ p $ large enough. One such \replaced{RSF}{example} is  $[0,1,2,4,6]$, and the resulting SR SCB codes \deleted[RW]{can} avoid\added[RW]{s} $ (4,8) $ absorbing sets for prime $p>23$.
\end{remark}

\subsection{$ (5,9) $ absorbing sets} \label{subsection59}
\replaced{Assuming an RSF that avoids the $(4,8)$ absorbing sets, this section proves that the $(5,9)$ absorbing sets are the smallest remaining.}{
Assuming $(4,8)$ absorbing sets do not exist in the SR SCB codes with \added[LD]{a} carefully chosen RSF, this section first proves the $(5,9)$ absorbing set is the second smallest absorbing set.} \added[LD]{The CCM approach shows that the $(5,9)$ absorbing sets always exist in the EAB SCB codes, but are avoided for SR SCB codes by some of the RCFs that precluded the $(4,8)$ absorbing sets.} \deleted[RW]{do not exist in SR SCB codes with a proper choice of RSF.}
\begin{lemma}\label{lemma5b}
Assuming $ (4,8) $ absorbing sets do not exist, $(5,b)$ absorbing sets also do not exist for $ b<9 $.
\end{lemma}

\textit{Proof:}
Appendix~\ref{pr:lemma5b} provides the proof.
\hfill$\blacksquare$

\begin{corollary}\label{cor5b}
Assuming $ (4,8) $ absorbing sets do not exist, the $(5,9)$ absorbing set is the smallest absorbing set.
\end{corollary}

\textit{Proof:}
Appendix~\ref{pr:cor5b} provides the proof.
\hfill$\blacksquare$

\begin{lemma}\label{lemma59}
$ \det \mathbf{M}=0 \mod p$ is necessary and sufficient for the existence of $ (5,9) $ absorbing sets in Fig.~\ref{fig59}.
\end{lemma}


\textit{Proof:}
Appendix~\ref{pr:lemma59} provides the proof.
\hfill$\blacksquare$

\added[RW]{Appendix~\ref{pr:lemma59num} shows that the number of $ (5,9) $ absorbing sets in the EAB codes scales as $\Theta(p^3)$.}
However, with \added[LD]{a} proper choice of RSF, $ \det \mathbf{M} \neq 0 \mod p $ for $ p $ large enough. Therefore we can conclude with \added[LD]{the} following two corollaries.
\begin{corollary} \label{cor:eab59}
$(5,9)$ absorbing sets exist in EAB codes described by the parity check matrix $H_{p,i \cdot j}^{5,p}$, and their number scales as $\Theta(p^3)$.
\end{corollary}
\begin{corollary} \label{cor:sr59}
There are no $(5,9)$ absorbing sets in the SR SCB codes described by the parity check matrix $H_{p, a(i)\cdot j}^{5,p}$, for prime $p$ large enough with \added[LD]{a}  proper choice of RSF.
\end{corollary}

\textit{Proof:}
For SR SCB codes, it is sufficient to select RSF such that~\eqref{eq59m} in Appendix \ref{pr:lemma59} does not evaluate to zero. One such example is $[0,1,2,4,7]$, where it is sufficient for the prime \added[JW]{$p>89$}  \added[LD]{and not be in the}  set $\{ 101,103,131,179\}$.  Therefore in the SR SCB code there is no $(5,9)$ absorbing set \added[RW]{if $p$ is sufficiently large and an appropriate RCF is chosen.}
\hfill$\blacksquare$

\subsection{$ (6,8) $ absorbing sets} \label{subsection68}
\replaced[RW]{This section considers the $(6,8)$ absorbing sets, which are the smallest remaining after the $(4,8)$ and $(5,9)$ absorbing sets.  We will investigate the $(6,8)$ absorbing sets both for EAB codes and for SR-SCB codes that preclude the $(4,8)$ and $(5,9)$ absorbing sets. For the six candidate configurations of $(6,8)$ absorbing sets, all satisfied checks have degree 2 and all unsatisfied checks have degree 1 in the absorbing set graph. Combinatorial and consistency arguments show that four of these six configurations are not present for $p$ sufficiently large in either the EAB code or in SR-SCB codes that preclude the $(4,8)$ and $(5,9)$ absorbing sets.  The remaining two configurations \deleted[JW]{shown in Fig.~\ref{fig64}} have the cardinality $\Theta(p^3)$ in the EAB code. However, both of these configurations contain a $(4,8)$ absorbing set as a subset and thus cannot be present in SR-SCB codes that preclude the $(4,8)$ \deleted[JW]{and $(5,9)$} absorbing sets.}
{Assuming $ (4,8) $ and $ (5,9) $ absorbing sets do not exist in the SR SCB codes with \added[LD]{a} carefully chosen RSF, this section considers the $(6,8)$ absorbing set, which is the third smallest absorbing set in the $H_{p,i \cdot j}^{5,p}$ code family. There are six candidate configurations of $(6,8)$ absorbing sets. In these non-isomorphic candidate $(6,8)$ absorbing sets, all satisfied checks (unsatisfied checks) incident to \replaced[JW]{variable}{bit} nodes in the absorbing set have degree 2 (degree 1). Out of these six configurations, we will show that four can be eliminated for $p$ large enough using combinatorial arguments and the bit consistency and check consistency conditions applied to either choice of $f(i,j)$. The remaining two configurations have the cardinality $\Theta(p^3)$ in the EAB SCB code, and interestingly, both  contain a $(4,8)$ absorbing set as a subset. These are therefore completely eliminated by the transformation on $f(i,j)$.}

\subsubsection{$ (6,8) $ configuration 1 - check nodes with degree$>$2}

\begin{lemma}\label{lemma3a}
For $H_{p,i \cdot j}^{5,p}$ and $p$ sufficiently large, there are no $(6,8)$ absorbing sets for which a check node connects to more than two \replaced[JW]{variable}{bit} nodes in the absorbing set graph.
\end{lemma}

\textit{Proof:}
Appendix~\ref{pr:lemma3a} provides the proof.
\hfill$\blacksquare$

Therefore attention is restricted to the case where all check nodes in the absorbing set graph have degree at most 2.

\begin{figure}\vspace{-0.2in}
\center
\includegraphics[width=0.2\textwidth]{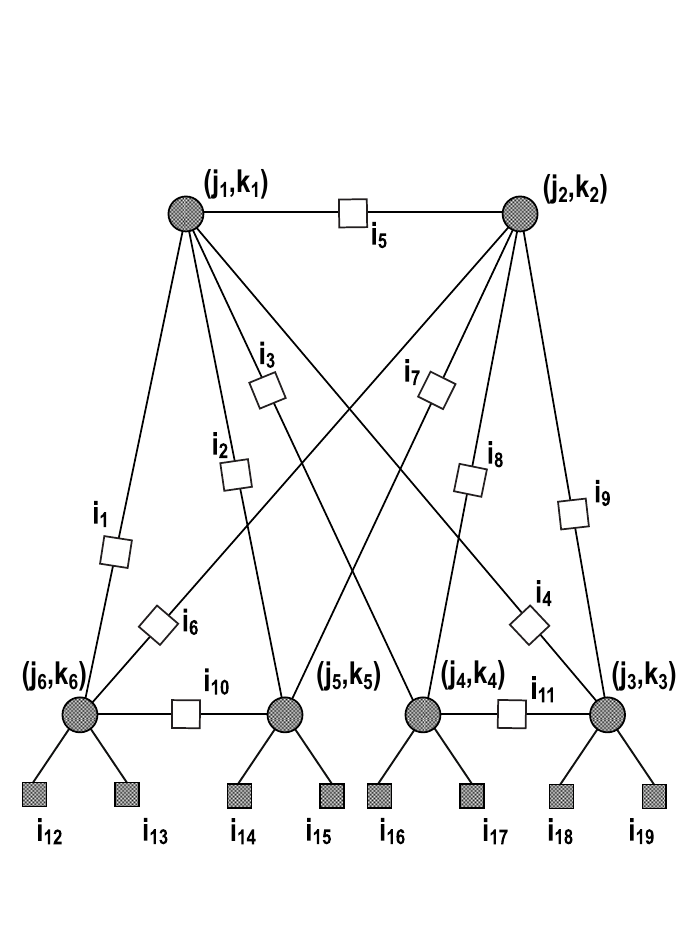}\vspace{-0.2in}
\caption{$ (6,8) $ configuration candidate 2.}\label{cand2}\vspace{-0.3in}
\end{figure}

In a candidate $(6,8)$ absorbing set,  \replaced[JW]{variable}{bit} nodes can have 3, 4 or 5 satisfied checks. By the girth constraint, there can be at most 2 \replaced[JW]{variable}{bit} nodes with 5 satisfied checks. Suppose there are two such \replaced[JW]{variable}{bit} nodes. Since there are a total of 8 unsatisfied checks, the other 4 \replaced[JW]{variable}{bit} nodes must each have 3 satisfied and 2 unsatisfied checks. This necessarily implies the configuration shown in Fig.~\ref{cand2} which we discuss next.
\subsubsection{$ (6,8) $ configuration candidate 2 - Fig.~\ref{cand2}}

Since the configuration in Fig.~\ref{cand2} \added[LD]{has two overlapping}  $ (4,8) $ absorbing sets, the existence of this $ (6,8) $ absorbing set relies on both CCM determinants of these two $ (4,8) $ absorbing sets that share an edge. Therefore with \added[LD]{a proof similar to that of Lemma~\ref{lemma48}, we can show the following result}:
\begin{lemma} \label{lemmacand2}
The necessary and sufficient condition for the existence of $ (6,8) $ absorbing sets in Fig.~\ref{cand2} is $ \det \mathbf{M}_1=0 \mod p$ and $ \det \mathbf{M}_2=0 \mod p$, where $ \mathbf{M}_1 $ and $ \mathbf{M}_2 $ are CCMs of the two internal $ (4,8) $ absorbing sets.
\end{lemma}

We can also prove that there are $p^2(p-1)$ such absorbing sets in the EAB codes with any prime $ p $. The proof is shown in Appendix~\ref{pr:lemmacand2}.

Similarly, the following corollaries are consequences of Corollary ~\ref{cor:eab48} and ~\ref{cor:sr48}.
\begin{corollary} \label{cor:eab68cand2}
The $(6,8)$ absorbing sets in Fig.~\ref{cand2} exist in EAB codes described by the parity check matrix $H_{p,i \cdot j}^{5,p}$, and their number scales as $\Theta(p^3)$.
\end{corollary}
\begin{corollary} \label{cor:sr68cand2}
There are no $(6,8)$ absorbing sets in Fig.~\ref{cand2} in the SR SCB codes described by the parity check matrix $H_{p, a(i)\cdot j}^{5,p}$, for prime $p$ large enough with proper choice of RSF.
\end{corollary}

Suppose now that there is exactly one \replaced[JW]{variable}{bit} node in the absorbing set having all five checks satisfied. The \replaced[JW]{variable}{bit} nodes in the absorbing set must necessarily be arranged either as in Fig.~\ref{figa3} or Fig.~\ref{figa4}.

\added[JW]{Similar analysis of $ (6,8) $ configuration candidates 3 to 6 is shown in Appendix~\ref{app:cand3}, ~\ref{app:cand4}, ~\ref{app:cand5} and \ref{app:cand6} respectively.}

\vspace{-0.1in}\subsection{Non-existence of $ (4,8) $, $ (5,9) $ or $ (6,8) $ absorbing sets in SR SCB codes \added[RW]{with a well-chosen RSF}} \label{nonexistence3sets}
The following is a consequence of Lemma~\ref{lemma48} to Lemma~\ref{lemma1a}, and Corollary~\ref{cor:sr48} to Corollary~\ref{coro1b}.
\begin{theorem}\label{thm_r5}
In  the EAB SCB code the number of $(4,8)$, $(5,9)$ and $(6,8)$ absorbing sets scales as $\Theta(p^3)$ whereas in an SR SCB code \added[RW]{with a well-chosen RSF} there are no $(4,8)$, $(5,9)$ or $(6,8)$ absorbing sets for \replaced{sufficiently large}{large enough} $p$.
\end{theorem}

\begin{remark}
For small $ p $'s, $ (4,8) $, $ (5,9) $ and $ (6,8) $ absorbing sets cannot be eliminated simultaneously for the SR SCB codes. Since there are 3 equivalence classes of SR codes~\cite{JWANGICC11}, we only need to consider the RSF that contains 0 and 1. Here are a few good RSFs for $ p>61 $ that eliminate all $ (4,8) $, $ (5,9) $ and $ (6,8) $ absorbing sets: \added[JW]{(1) $[0,1,2,4,17]$ for $ p=67 $ (2) $[0,1,2,3,11]$ for $ p=73 $ (3) $[0,1,2,6,7]$ for $ p=79 $ (4) $[0,1,2,3,7]$ for $ p=83,97,101,103,107,109,113,127$ (5) $[0,1,2,4,11]$ for $ p=89 $ (6) $ [0,1,2,4,7] $ for $ p>179 $.}
\end{remark}

\vspace{-0.1in}\subsection{Absorbing set spectrum in the Tanner construction}\label{tannerfor5}
We can easily extend our analysis \deleted[ZZ]{and apply it} to the Tanner construction in \cite{tanner04}, which is an example of SSR SCB codes.
\begin{lemma}\label{lemmaqc1}
In the Tanner graph corresponding to quasi-cyclic LDPC $H_{p, \tilde{f}(i,j)}^{5,p}$ in \cite{tanner04}, no $ (4,8) $ or $(6,8)$ absorbing set exists with parameters selected in \replaced[RW]{Table}{TABLE} I of \cite{tanner04}.
\end{lemma}

\textit{Proof:}
Proof is shown in Appendix~\ref{pr:lemmaqc1}.
\hfill$\blacksquare$

\begin{remark}
\replaced{This}{Tanner's} code \added[LD]{has a} good absorbing set spectrum and thus will have low error floor as expected. Moreover, the analysis can be easily extended to the quasi-cyclic codes constructed in \cite{FossorierIT04}, which have the first sub-row and first sub-column as identity matrices.
\end{remark}

\added[LD]{Note that the codes listed in \replaced[RW]{Table}{TABLE} I of \cite{tanner04} are mostly \replaced{moderate-rate}{moderate rates} codes. }However, for higher rates, the \replaced{Tanner construction}{structure of Tanner's codes} \added[LD]{may}  introduce smaller absorbing sets~\cite{WangITA2011}. These absorbing sets can also be avoided by a carefully chosen CSF that precludes the $ \mathbf{u} $ from the null space of $ \mathbf{M} $.

\vspace{-0.05in}
\section{Results}\label{results}


In this section we experimentally demonstrate performance improvement with well-designed SR codes. In simulations, we use 200 iterations and a $ Q4.2 $ fixed-point quantization with 4, resp. 2, bits to represent integer, resp. fractional, values. We simulate sum-product algorithm~\cite{zhangTCOM} on an FPGA platform.

\begin{figure}[!t]
\center
\includegraphics[width=0.4\textwidth]{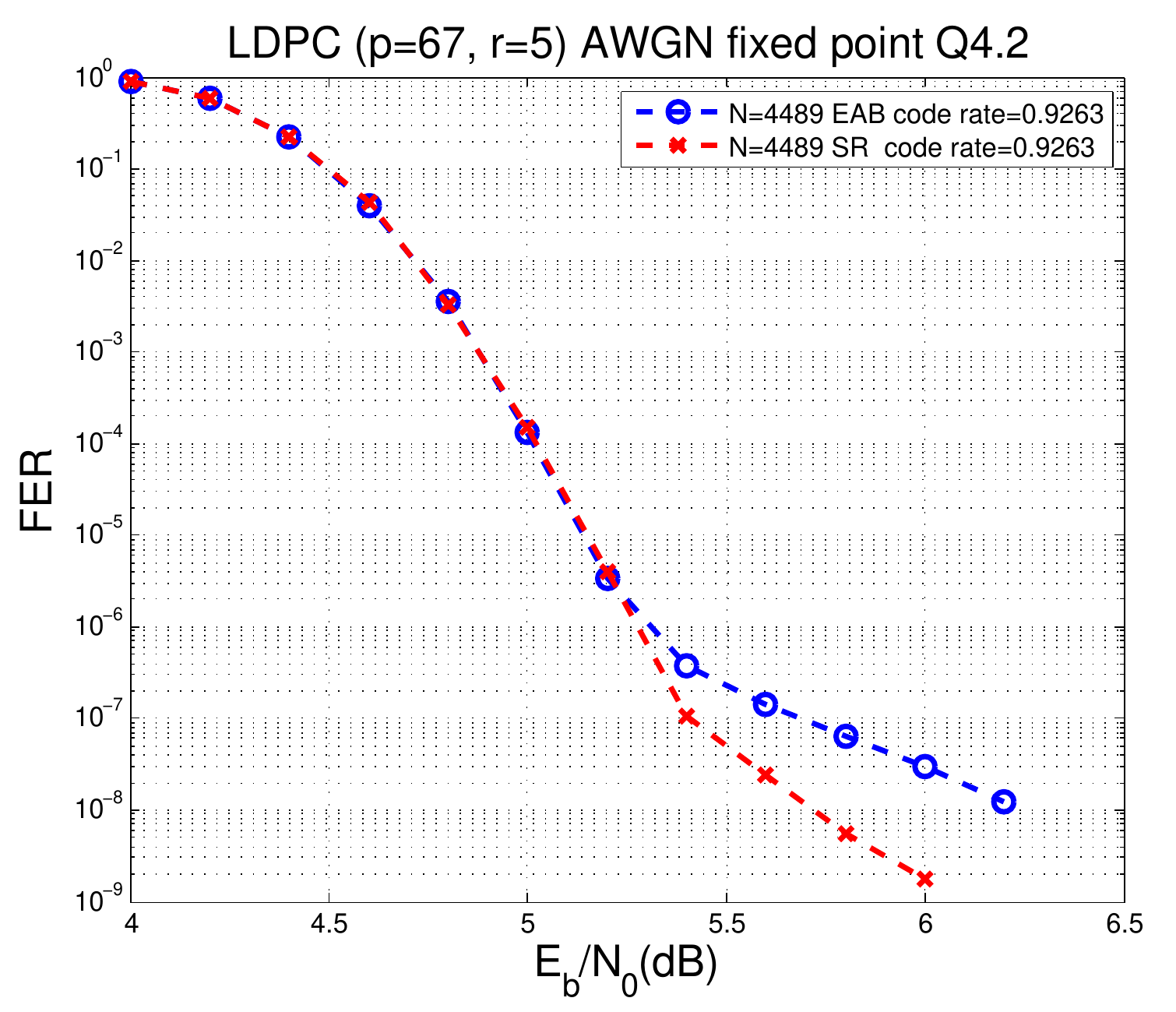}
\caption{Performance comparison of the $ (4489,4158) $ EAB SCB and SR SCB LDPC codes. }\vspace{-0.2in}
\label{fig:perf4}
\end{figure}

We simulated a pair of longer block length codes contrasted in Fig.~\ref{fig:perf4}. The performance improvement of the SR-SCB code is \added[LD]{due to}  the elimination of the $ (4,8) $, $ (5,9) $ and $ (6,8) $ absorbing sets with \added[LD]{a}  proper choice of the \added[JW]{RSF}. (Here the SR-SCB code's \replaced[JW]{RSF is $ [0,1,2,4,17] $}{$(i,a(i))$ $\in$ $\{(0,0), (1,1), (2, 2), (3,4), (4, 17)\}$}.) We \deleted[JW]{can} observe that in the error profile as shown in Table~\ref{tab_fpga_new4k} the $ (4,8) $, $ (5,9) $ and $ (6,8) $ absorbing sets are completely eliminated in the SR SCB code.  

Row selection alone cannot avoid the next smallest absorbing sets, which are the $(6,10)$.  Each of these $(6,10)$ absorbing sets exists if and only if the corresponding subset $(6,4)$ absorbing set studied in~\cite{JWANGICC11} exists.  As shown in \cite{JWANGICC11}, these $(6,4)$ absorbing sets cannot be eliminated only with row selection.  However, \cite{JWANGICC11} also shows that these absorbing sets can be precluded by column selection.

\small
\begin{table}
\caption{\deleted[ZZ]{Hardware}Error profiles for the EAB SCB $(4489,4158)$, code (top), and the SR SCB code (bottom). n.e. is the number of collected errors.}\label{tab_fpga_new4k}
\vspace{-.1in}
\begin{center}
\begin{tabular}{|p{0.18in}|p{0.15in}|p{0.14in}|p{0.14in}|p{0.14in}|p{0.18in}|p{0.14in}|p{0.18in}|p{0.14in}|p{0.14in}|p{0.18in}|}
\hline
SNR & n.e. &(4,8)&(5,9)&(6,8)&(6,10)&(7,9)&(7,11)&(8,6)&(8,8)&(8,10)\\
\hline
5.6dB& 150 & 67&17&22&7&6&5&6&6&3\\
5.8dB& 139 & 83&18&16&6&5&1&3&3&1\\
6.0dB& 131 & 77&18&22&5&1&1&2&1&1\\
6.2dB& 107 & 85&10&5&4&2&0&0&0&0\\
\hline
\hline
\end{tabular}

\begin{tabular}{|p{0.18in}|p{0.15in}|p{0.14in}|p{0.14in}|p{0.14in}|p{0.18in}|p{0.14in}|p{0.18in}|p{0.14in}|p{0.14in}|p{0.18in}|}
\hline
SNR & n.e. &(4,8)&(5,9)&(6,8)&(6,10)&(7,9)&(7,11)&(8,6)&(8,8)&(8,10)\\
\hline
5.6dB& 106 & 0&0&0&25&15&6&15&13&6\\
5.8dB& 140 & 0&0&0&35&29&14&16&6&8\\
6.0dB& 60 & 0&0&0&25&7&5&9&5&3\\
\hline
\end{tabular}
\end{center}
\vspace{-0.3in}
\end{table}
\normalsize

\section{Conclusion}\label{conclusion}\vspace{-0.02in}
\added[LD]{This paper presents a detailed analysis of the absorbing set spectrum of a class of LDPC codes based on circulant matrices. Using the cycle consistency matrix description of the dominant absorbing sets we characterized code performance and provided tools for a systematic code design. \deleted[ZZ]{Hardware}\added[ZZ]{Simulation} results in low FER region support the proposed methodology.}
\bibliographystyle{unsrt}	
\vspace{-0.2in}\bibliography{myrefs}		

\section{Appendix}
\subsection{Proof of Theorem~\ref{thm:ccm}}\label{pr:thmccm}
Each of the three conditions has already been shown to be a necessary condition for the existence of $G_\text{as}$ in an SCB. If all of these three conditions are satisfied, all the cycles presented in the CCM exist in $G_H$ and any linear combination of these cycles exists in $G_H$ as well.  The only issue is whether the existing graphical structures have {\em additional} linearly independent cycles not required by the CCM. There are only three ways for this to happen: (1) a variable node's unsatisfied check node is the same as another variable node's unsatisfied check node, or (2) a variable  node's unsatisfied check node is the same as one satisfied check node in the graph, or (3) two of the satisfied check nodes are the same.  In each of these cases, additional edges extend the VN graph.  For this to be possible, the original VN graph must be extensible as defined above. Thus if the VN graph is not extensible, the above constructed solution fully describes the existence of the proposed absorbing set. This concludes that $G_{\text{as}}$ is present in $G_H$.
\subsection{Proof of Corollary~\ref{cor:ext}} \label{pr:corext}
Suppose the CCMs of $G^{1}_{\text{as}}$ and $G^{2}_{\text{as}}$ are $ \mathbf{M}_1 $ and $ \mathbf{M}_2 $ respectively. If the VN graph of $G^{1}_{\text{as}}$  is a sub-graph of the VN graph of $G^{2}_{\text{as}}$, the independent cycles of $G^{1}_{\text{as}}$ will also be independent in $G^{2}_{\text{as}}$ and thus $ \mathbf{M}_1 $ could be a sub-matrix of $ \mathbf{M}_2 $:
\small
\begin{equation}
\mathbf{M}_2 = \left[
\begin{array}{cc}
\mathbf{M}_1 & 0\\
\mathbf{A} & \mathbf{B}
\end{array}
\right].
\end{equation}
\normalsize
Therefore if there exists a valid $ \mathbf{u}_2 $ such that $ \mathbf{M}_2 \mathbf{u}_2 = 0 \mod p$, the first $ a_1 $ elements would also be a valid $ \mathbf{u}_1 $ such that $ \mathbf{M}_1 \mathbf{u}_1 = 0 \mod p$. This concludes that the existence of $G^{1}_{\text{as}}$ is a necessary condition of the existence of $G^{2}_{\text{as}}$.
\subsection{Proof of Lemma~\ref{lemma48}} \label{pr:lemma48}
Since the VN graph of the $ (4,8) $ absorbing sets is a fully connected graph, it is not extensible without introducing a length-4 cycle in the corresponding bipartite graph. According to Theorem~\ref{thm:ccm}, $ \det \mathbf{M}=0 \mod p$ is a necessary condition since it implies the $ \mathbf{M} $ is not full column-rank, and it is a sufficient condition if bit consistency and check consistency are both satisfied. We can verify both of these conditions by carefully constructing a solution in the null space of $ \mathbf{M} $.

Since the binary cycle space for Fig.~\ref{fig48} has dimension $3$, we construct the following CCM by selecting the following 3 linearly independent cycles: {$v_1-v_2-v_3,v_1-v_2-v_4,v_1-v_3-v_4$}:
\small
\begin{equation}\label{eq48}
\mathbf{M} =\left[
\begin{array}{ccc}
i_1-i_2 & i_2-i_5 & 0\\
i_1-i_6 & 0 & i_6-i_4 \\
0 & i_5-i_3 & i_3-i_4
\end{array}
\right].
\end{equation}
\normalsize
If $ \det{\mathbf{M}} \equiv 0 \mod p $, there exists a non-zero solution to $ \mathbf{M} \cdot \mathbf{u} \equiv 0 \mod p$, where $ \mathbf{u}=[u_2,u_3,u_4]^T $. Without loss of generality, suppose $ u_2 \neq 0 $. With check consistency, $ i_1-i_2 \neq 0 $,$ i_2-i_5 \neq 0 $, and $ i_1-i_2 \neq i_2-i_5 $. Thus $ u_3 \neq 0 $ and $ u_2 \neq u_3 $. Similarly $ u_4 \neq 0 $, $ u_2 \neq u_4 $, and $ u_3 \neq u_4 $. Then, for a fixed $ j_1 $, we can find $ j_2 $, $ j_3 $, and $ j_4 $ without contradiction to the bit consistency. With any specific $ k_1 $ values, we can find $ (4,8) $ absorbing sets in the code. Therefore $ \det{\mathbf{M}} \equiv 0 \mod p $ is a sufficient condition for the existence of $ (4,8) $ absorbing sets.
\subsection{Proof of Corollary~\ref{cor:eab48} and Corollary~\ref{cor:sr48} } \label{pr:cor2and3}
From the proof in Appendix~\ref{pr:lemma48}, $ \det{\mathbf{M}} \equiv 0 \mod p $ is a necessary and sufficient condition for the existence of $ (4,8) $ absorbing sets.

Under the bit consistency and check consistency, there are only two possible non-isomorphic check labelings \cite{DOLECEKISTC10}: $(i_1,i_2,i_3,i_4,i_5,i_6)$ to be either $(x,y,x,y,z,w)$ for  $\{ x, y, z, w \} \subset \{0, 1, 2, 3, 4\}$  (assignment 1) or $(x,t,w,y,z,z)$  (assignment 2). Thus $ \det \mathbf{M} \equiv 0 \mod p $ implies the following necessary and sufficient conditions for the existence of a $(4,8)$ absorbing set result:
\begin{equation}\label{eq11}
(z - x)(w - y) + (z - y)(w - x) \equiv 0 \mod p~,
\end{equation}
for assignment 1, and
 \begin{equation}\label{eq2}
(z-w)(x-t)(y-z)-(y-w)(x-z)(z-t) \equiv 0 \mod p~,
\end{equation}
for assignment 2.

For EAB codes $\{x, y, z, w,t\} = \{0, 1, 2, 3, 4\}$, in the former case, in fact one can show there are no solution sets for prime $p$ large enough ($p > 17$).
For the latter case there are $8$ solution sets	$(x, y, z, w, t) \in$ $\left\{(4,3,2,0,1),(4,1,2,0,3),(3,4,2,1,0),(3,0,2,1,4), \right.$
$\left. (1,4,2,3,0),(1,0,2,3,4),(0,3,2,4,1),(0,1,2,4,3)\right\}$ that always evaluate to zero on the left-hand side of equation~\eqref{eq2}. These numerical solutions are in fact symmetric so that once the labels of the check nodes are selected (cf. Fig.~\ref{fig48}), the \replaced[JW]{variable}{bit} node labels (pairs $(j_1,k_1)$ through $(j_4,k_4)$) can be selected in $\Theta(p^3)$ ways, thereby completely characterizing the absorbing set of interest.

\added[LD]{We say that an absorbing set is fully absorbing set if, in addition, all \replaced[JW]{variable}{bit} nodes \textit{outside} the absorbing set have more satisfied than unsatisfied checks.} 

\deleted[LD]{Moreover, such an} \added[LD]{Then the above} $(4,8)$ absorbing set is always a $(4,8)$ fully absorbing set since otherwise there would exist a \replaced[JW]{variable}{bit} node $(j_5,k_5)$ outside the absorbing set incident to at least three of the checks labeled $i_7$ through $i_{14}$. Such a configuration would either violate the girth  constraint, or it would imply the existence of a new configuration spanning four \replaced[JW]{variable}{bit} nodes (these being the node $(j_5,k_5)$  and three \replaced[JW]{variable}{bit} nodes from the starting $(4,8)$ absorbing set). These four \replaced[JW]{variable}{bit} nodes would necessarily be connected such that their  common constraint is given in~\eqref{eq11}, previously shown to not hold for large enough $p$.

To avoid this absorbing set we need to force the determinant of the CCM to be nonzero. It is sufficient to assign values to $a(i)$ such that the selected labels for the check nodes $i_1$ through $i_6$ do not satisfy equations~\eqref{eq11} and~\eqref{eq2}. One such example is  $(i,a(i))$ $\in$  $\{(0,0), (1,1), (2, 2), (3,4), (4, 6)\}$.  Therefore in the SR SCB code there is no $(4,8)$ absorbing set for large enough $p$.
\subsection{Proof of Lemma~\ref{lemma5b}}\label{pr:lemma5b}
Since the total edge number is odd and the number of edges that go to satisfied check nodes is even, $ b $ can only be odd. Thus if $ b<9 $, $ b $ can only choose values from $ \{1,3,5,7\} $. For these $ b $ values, the corresponding VN graph will always contain the VN graph of $ (4,8) $ absorbing sets without introducing a length-4 cycle in the bipartite graph. By Corollary~\ref{cor:ext}, the $ (5,b) $ absorbing sets with $ b<9 $ do not exist if $ (4,8) $ absorbing sets are absent.

\subsection{Proof of \deleted[LD]{Lemma}\added[LD]{Corollary}~\ref{cor5b}}\label{pr:cor5b}
Since the number of edges that go to unsatisfied check nodes is at most 10 and $ b $ is odd, the only possible $ (5,b) $ absorbing set is $ (5,9) $ absorbing set in the absence of $ (4,8) $ absorbing sets. Since the only possible configuration of a $ (5,9) $ absorbing set is one variable node that has $ 4 $ satisfied check nodes and four variable nodes that have $ 3 $ satisfied check nodes. Fig.~\ref{fig59} depicts this configuration, which has a VN graph that does not contain the VN graph of $ (4,8) $ as a sub-graph. Therefore we can use \deleted[LD]{the following lemma}\added[LD]{Lemma~\ref{lemma59}} to build the necessary and sufficient condition of this absorbing set.
\subsection{Proof of Lemma~\ref{lemma59}}\label{pr:lemma59}
The binary cycle space for Fig.~\ref{fig59} has dimension $4$. We construct the following CCM by selecting the following linearly independent cycles: {$v_1-v_2-v_3,v_1-v_2-v_5,v_1-v_3-v_4,v_1-v_4-v_5$}:
\small
\begin{equation}\label{eq59}
\mathbf{M} =\left[
\begin{array}{cccc}
i_1-i_5 & i_5-i_2 & 0 & 0\\
i_1-i_8 & 0 & 0 & i_8-i_4 \\
0 & i_2-i_6 & i_6-i_3\\
0 & 0 & i_3-i_7 & i_7-i_4
\end{array}
\right].
\end{equation}
\normalsize
Similarly to the proof of Lemma~\ref{lemma48}, we can show that $ \det \mathbf{M} = 0 \mod p $ implies the following necessary and sufficient conditions for the existence of a $(5,9)$ absorbing set, where
\small
\begin{equation} \label{eq59m}
\begin{split}
\det \mathbf{M} =& (i_1-i_5)(i_8-i_4)(i_2-i_6)(i_3-i_7)\\
&-(i_1-i_8)(i_5-i_2)(i_6-i_3)(i_7-i_4) \mod p
\end{split}
\end{equation}
\normalsize
\subsection{Number of $ (5,9) $ absorbing sets in EAB codes}\label{pr:lemma59num}
Under the bit consistency and check consistency, there are $ 5 $ possible non-isomorphic check labelings: $(i_1,i_2,i_3,i_4,i_5,i_6,i_7,i_8)$ to be  $(x,y,z,w,z,w,x,y)$, $(x,y,z,w,t,w,x,y)$, $(x,y,z,w,t,w,x,z)$, $(x,y,z,w,t,x,t,y)$, or $(x,y,z,w,t,w,t,y)$. The EAB codes have  $\{ x, y, z, w,t \}$ $ \subset$ $ \{0, 1, 2, 3, 4\}$. For the $ 4 $th case, there are $8$	solution sets $(x, y, z, w, t)$ $ \in$ $\left\{(4,0,1,3,2),(4,0,3,1,2),(3,1,4,0,2),(3,1,0,4,2), \right.$ $\left. (1,3,4,0,2),(1,3,0,4,2),(0,4,1,3,2),(0,4,3,1,2)\right\}$ that always evaluate to zero on the left-hand side of equation~\eqref{eq59m}. Once the labels of the check nodes are selected (cf. Fig.~\ref{fig59}), the variable node labels (pairs $(j_1,k_1)$ through $(j_5,k_5)$) can be selected in $\Theta(p^3)$ ways, thereby completely characterizing the absorbing set of interest.
\subsection{Proof of Lemma~\ref{lemma3a}}\label{pr:lemma3a}
\begin{figure}
\centering
\includegraphics[width=0.25\textwidth]{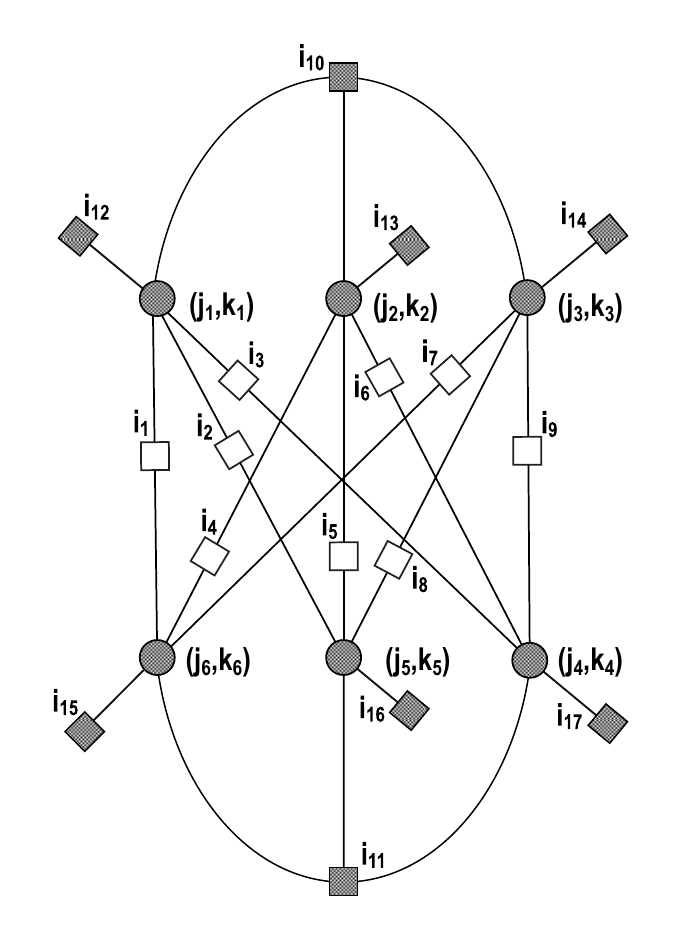}\vspace{-0.2in}
\caption{$ (6,8) $ configuration candidate 1.}\label{cand1}\vspace{-0.2in}
\end{figure}
For those $ (6,8) $ absorbing sets with the property that a check node connects to more than 2 \replaced[JW]{variable}{bit} nodes in the absorbing set, the corresponding VN graph will always contain the VN graph of $ (4,8) $ absorbing sets without introducing a length-4 cycle in the bipartite graph. One such example is shown in Fig.~\ref{cand1}. Since this structure contains several $ (4,8) $ absorbing sets, we establish a CCM for each $ (4,8) $ absorbing set. \deleted[LD]{If this configuration in Fig..~\ref{cand1} exist, the CCMs of all inside $ (4,8) $ absorbing sets need to have zero determinant, which cannot be satisfied simultaneously for $ p $ large enough.} \added[LD]{If the configuration in Fig.~\ref{cand1} were to exist, the CCMs of all  $ (4,8) $ absorbing sets within this configuration would  have zero determinant, resulting in a set of conditions which cannot be satisfied simultaneously for $ p $ large enough.}
\subsection{Proof of Lemma~\ref{lemmacand2}}\label{pr:lemmacand2}
\begin{figure}
\center
\includegraphics[width=0.25\textwidth]{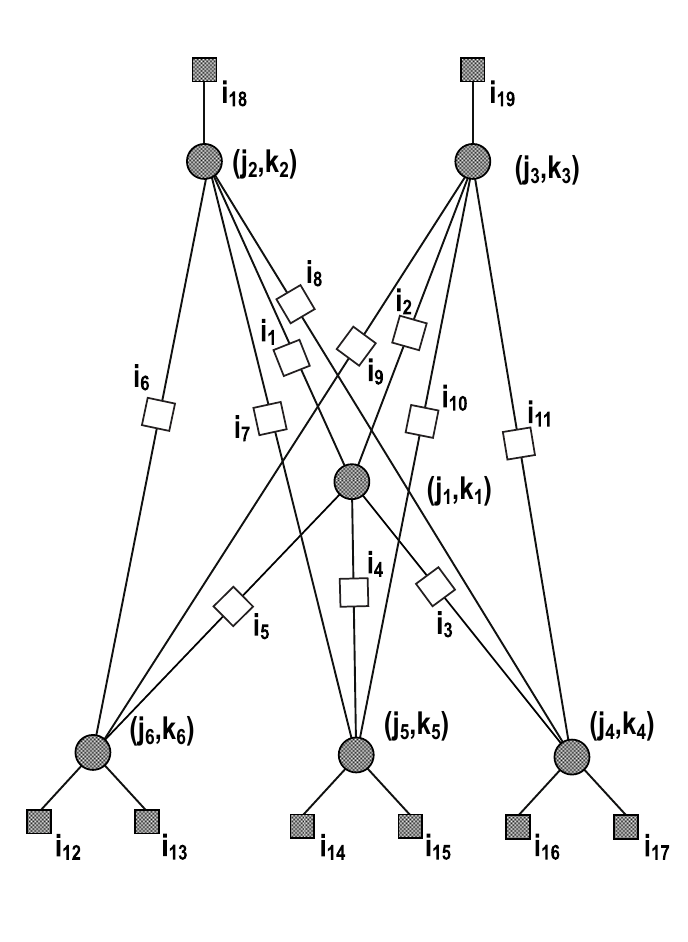}\vspace{-0.2in}
\caption{$ (6,8) $ configuration candidate 3.}\vspace{-0.2in}
\label{figa3}
\end{figure}

We may start with the substructure spanning 4 \replaced[JW]{variable}{bit} nodes $(j_1,k_1)$, $(j_2,k_2)$, $(j_5,k_5)$ and $(j_6,k_6)$.
Recall that out of six check nodes shared by these \replaced[JW]{variable}{bit} nodes, exactly two have the same label (it being label ``2"), and that there is $p^2(p-1)$ ways of assigning values of the \replaced[JW]{variable}{bit} nodes and the check nodes in this substructure. By symmetry of the configuration it suffices to consider the case when this repeated label is the one corresponding to $i_5$ and $i_{10}$ and when this repeated label corresponds to some other pair of parallel edges. The latter case is not possible since the \replaced[JW]{variable}{bit} nodes  $(j_1,k_1)$, $(j_2,k_2)$, $(j_3,k_3)$ and $(j_4,k_4)$ themselves constitute a $(4,8)$ absorbing set and there the propagation of the check labels through the proposed configuration necessarily violates  the check label constraints. In the former case, by the bit consistency constraints, the labeling of the checks incident to $(j_3,k_3)$ and $(j_4,k_4)$ is unique for each of the nodes (without assuming these two nodes themselves share an edge). Moreover, for the independently selected values of $(j_3,k_3)$ and $(j_4,k_4)$, we show by the pattern consistency constraint that they indeed have a common check $i_{10}$,  itself labeled ``2". As a result, once the values  $(j_1,k_1)$, $(j_2,k_2)$, $(j_5,k_5)$ and $(j_6,k_6)$ and their shared checks are pinned down -- which can be done in $p^2(p-1)$ ways -- the rest of the proposed $(6,8)$ configuration follows uniquely. As a result, the cardinality of $(6,8)$ absorbing sets of the type in Fig.~\ref{cand2} is \deleted[LD]{$p^2(p-1)$}\added[LD]{ $p^2(p-1)$}.
\subsection{$ (6,8) $ configuration candidate 3 - Fig.~\ref{figa3}}\label{app:cand3}

\begin{lemma}\label{lemma2a}
In the Tanner graph corresponding to $H_{p, f(i,j)}^{5,p}$ for the EAB SCB and for the SR SCB code there are no $(6,8)$ absorbing sets for $p$ large enough of the type shown in Fig.~\ref{figa3}.
\end{lemma}

\textit{Proof:}
Without loss of generality we may assign check node labels for the checks emanating from the \replaced[JW]{variable}{bit} node $(j_3,k_3)$ as follows:
$i_1=x$,$i_2=y$, $i_{3}=z$, $i_{4}=w$, and $i_5=t$, where $x,y,z,w,t$ are the five distinct check labels.

The binary cycle space for Fig.~\ref{figa3} has dimension $6$. We construct the following CCM by selecting the following linearly independent cycles: {$v_1-v_2-v_3,v_1-v_3-v_6,v_1-v_2-v_5,v_1-v_3-v_4,v_1-v_3-v_5,v_1-v_2-v_4$}:
\begin{equation}\label{eq2a000}
\mathbf{M} =\left[
\begin{array}{cccccc}
x-i_6 & 0 & 0 & 0 & i_6-t\\
0 & y-i_9 & 0 & 0 & i_9-t \\
x-i_7 & 0 & 0 & i_7-w & 0 \\
0 & y-i_{11} & i_{11}-z & 0 & 0 \\
0 & y-i_{10} & 0 & i_{10}-w & 0 \\
x-i_8 & 0 & i_8-z & 0 & 0
\end{array}
\right].
\end{equation}
\normalsize
The rank of the matrix is at most 5 so if fact we may consider the top-left 5 by 5 submatrix (call it $B$). If the matrix $B$ is full rank, the only solution is $j_1 =j_2= j_3=j_4=j_5=j_6 $. Hence $\det(B)=0$ is necessary for the existence of the absorbing sets of this type.
Such condition can be expressed as
\begin{equation}\label{eq2a111}
\begin{split}
&-(i_{11}-z) [ -(x-i_6)(i_9-t)(i_7-w)(y-i_{10})\\
&+(x-i_7)(i_6-t)(y-i_9)(i_{10}-w)]\equiv 0 \mod p.
\end{split}
\end{equation}

Also consider the bottom-right 4 by 4 submatrix (call it $ A $). If the matrix $ A $ is full rank, the whole matrix will be rank 5, which only have one solution as $j_1 =j_2= j_3=j_4=j_5=j_6 $. Hence $\det(A)=0$ is necessary for the existence of the absorbing sets of this type.
Such condition can be expressed as
\begin{equation}\label{eq2a112}
\begin{split}
 -(x-i_7)(i_{10}-w)(y-i_{11})(i_{8}-z)\\
 +(x-i_8)(i_7-w)(i_{11}-z)(y-i_{10})\equiv 0 \mod p.
\end{split}
\end{equation}

For the values of $i_6,i_7,i_9,i_{10}$ and $i_{11}$ in the set described by $f(i,j)$ for both the EAB SCB and for the SR SCB codes, and such that the labels meeting at the same vertex are distinct (see Fig.~\ref{figa3}), the equation~\eqref{eq2a111} and ~\eqref{eq2a111} evaluate to zero for only finite number of values of the parameter $p$. For $a(i)=i$ (EAB SCB code) $\det(B) \neq 0$ for $p>23$. For the carefully designed SR SCB code, the equation~\eqref{eq2a111} and ~\eqref{eq2a111} also evaluate to zero for only finite number of values of the parameter $p$.
Thus for the EAB SCB code and for the SR SCB code there are no $(6,8)$ absorbing sets for $p$ large enough of the type shown in Fig.~\ref{figa3}.
\hfill$\blacksquare$

\begin{corollary}\label{corollary2a}
In the Tanner graph corresponding to $H_{p, f(i,j)}^{5,p}$, $(6,8)$ absorbing sets of the type shown in Fig.~\ref{figa3} exist if and only if $\det(A)=0$ and $\det(B)=0$.
\end{corollary}

\textit{Proof:}
Since the rank of a matrix is lower-bounded by the rank of its submatrix, and $ \mathbf{M} $ has a submatrix
\begin{equation}\label{eq2a001}
\left[
\begin{array}{cccc}
0 &  0 & 0 & i_9-t \\
x-i_7  & 0 & i_7-w & 0 \\
0  & i_{11}-z & 0 & 0 \\
0  & 0 & i_{10}-w & 0
\end{array}
\right],
\end{equation}
which is a full-rank 4 by 4 matrix, the rank of $ \mathbf{M} $ is no less than 4.

If absorbing sets of the type in Fig.~\ref{figa3} exist, $ \mathbf{M} $ is not full column-rank, otherwise the equation only has one solution as $j_1 =j_2= j_3=j_4=j_5=j_6 $. With the fact that $ \operatorname{rank}(\mathbf{M}) \geq 4 $ and $ \operatorname{rank}(\mathbf{M})<5 $, we can have $ \operatorname{rank}(\mathbf{M})=4 $, which implies $ \det(A)=0 $. (Otherwise $ \mathbf{M} $ will have rank 5 with the additional column.) Without loss of generality, we can consider the bottom row of $ A $ is dependent of the other 3 rows, and thus this row is redundant. Then $ \operatorname{rank}(B)=\operatorname{rank}(\mathbf{M})=4 $, which implies $ \det(B)=0 $.

If $ \det(A)=0 $ and $ \det(B)=0 $, without loss of generality, we consider the bottom row of $ A $ is dependent of the other 3 rows and thus $ \operatorname{rank}(B)=\operatorname{rank}(\mathbf{M}) $. Due to $ \det(B)=0 $, $ \mathbf{M} $ is not full rank and $ \operatorname{rank}(\mathbf{M})<5 $. With the fact that $ \operatorname{rank}(\mathbf{M}) \geq 4 $, we have $ \operatorname{rank}(\mathbf{M})=4 $ and there exist a non-zero $ \mathbf{u} $ such that equation~\eqref{eq2a000} is satisfied. Each element of $ \mathbf{u} $ is non-zero, otherwise the $ \mathbf{u} $ will be a zero vector.

This completes the proof of the corollary.
\hfill$\blacksquare$

\begin{figure}
\centering
\includegraphics[width=0.25\textwidth]{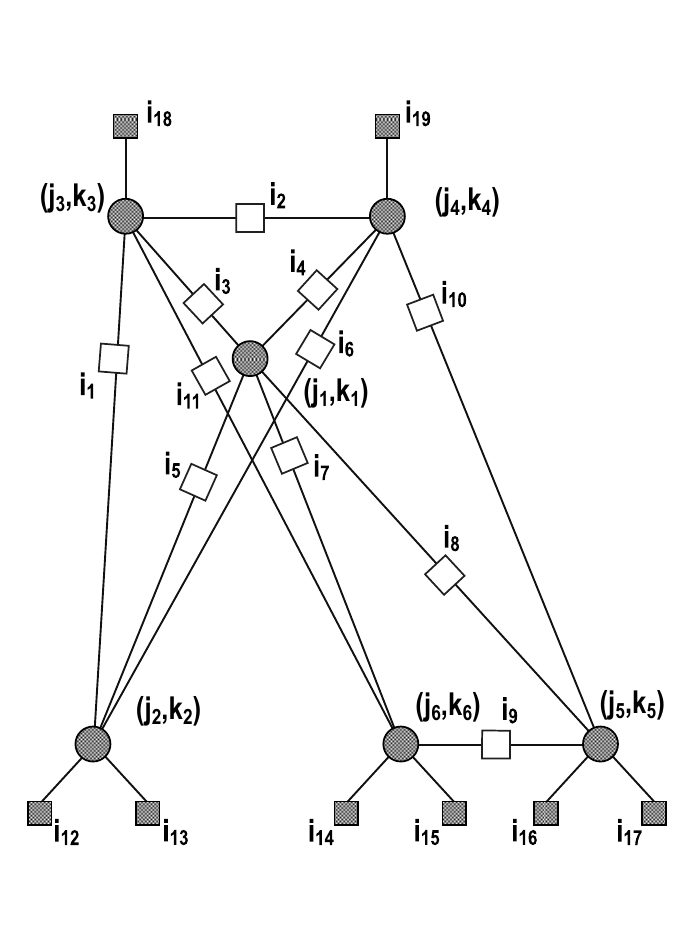}\vspace{-0.2in}
\caption{$ (6,8) $ configuration candidate 4.}\vspace{-0.2in}
\label{figa4}
\end{figure}

\subsection{$ (6,8) $ configuration candidate 4 - Fig.~\ref{figa4}}\label{app:cand4}

\begin{lemma}\label{lemmacand4}
In the EAB codes corresponding to $H_{p,i \cdot j}^{5,p}$, there are no $(6,8)$ absorbing sets with the topology shown in Fig.~\ref{figa4} for $p$ large enough.
\end{lemma}

\textit{Proof:}
The binary cycle space for Fig.~\ref{figa4} has dimension $6$. We construct the following CCM by selecting the following linearly independent cycles: {$v_1-v_2-v_3,v_1-v_2-v_4,v_1-v_3-v_4,v_1-v_3-v_6,v_1-v_4-v_5,v_1-v_5-v_6$}:
\begin{equation}\label{eq2a000}
\mathbf{M} =\left[
\begin{array}{cccccc}
i_5-i_1 & i_1-i_3 & 0 & 0 & 0\\
i_5-i_6 & 0 & i_6-i_4 & 0 & 0 \\
0 & i_3-i_2 & i_2-i_4 & 0 & 0 \\
0 & i_{11}-i_3 & 0 & 0 & i_7-i_{11} \\
0 & 0 & i_{10}-i_4 & i_{8}-i_{10} & 0 \\
0 & 0 & 0 & i_9-i_8 & i_7-i_9
\end{array}
\right].
\end{equation}
\normalsize
It suffices to consider the case when the labels $i_1,i_2,i_3,i_4,i_5,i_6$ adopt the  following assignment $(i_1,i_2,i_3,i_4,i_5,i_6)= (t,w,z,y,x,z)$ or when they are

 \noindent $(i_1,i_2,i_3,i_4,i_5,i_6)= (t,z,x,y,z,w)$

Analogously to proof of Lemma~\ref{lemma2a}), the $ \mathbf{M} $ is not full column-rank if $ \det A \neq 0 \mod p $ and $ \det B \neq 0 \mod p $, where $ A $ is the left-top 3 by 3 submatrix of $ \mathbf{M} $ and $ B $ is the right-bottom 4 by 4 submatrix of $ \mathbf{M} $.
This constraint cannot be satisfied for $p>41$ for $i_1$ to $i_{11}$ taking values in the set $\{ 0,1,2,3,4\}$ and such that the bit consistency constraints are satisfied for both labellings.
\hfill$\blacksquare$

Note that the variable-node of the configuration shown in Fig.~\ref{figa4} is a subgraph of the VN graph of $ (4,8) $ absorbing sets. The following again is an easy consequence of Corollary~\ref{cor:sr48}.
\begin{corollary}
If $ (4,8) $ absorbing sets are absent, the configuration shown in Fig.~\ref{figa4} is not possible in the SR SCB code.
\end{corollary}

\begin{figure}
\centering
\includegraphics[width=0.25\textwidth]{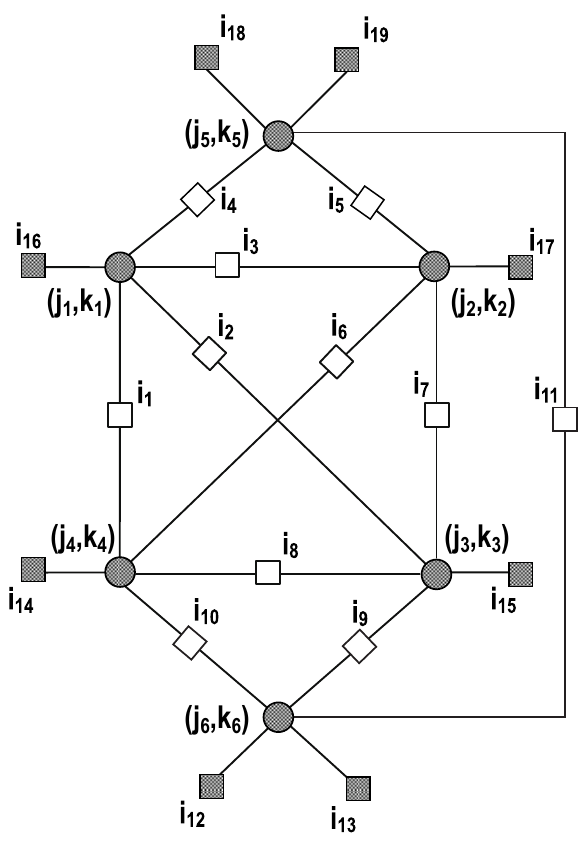}\vspace{-0.1in}
\centering
\caption{$ (6,8) $ configuration candidate 5.}\vspace{-0.2in}
\label{figa5}
\end{figure}

\subsection{$ (6,8) $ configuration candidate 5 - Fig.~\ref{figa5}}\label{app:cand5}
In the remainder we consider the case when no \replaced[JW]{variable}{bit} nodes in the absorbing set has all five satisfied checks. This constraint implies configurations shown in Fig.~\ref{figa5} and Fig.~\ref{figa6}.

\begin{lemma}\label{lemma1b}
In the EAB codes corresponding to $H_{p, i \cdot j }^{5,p}$ there are $\Theta(p^3)$  $(6,8)$ absorbing sets of the type shown in Fig.~\ref{figa5} for $p$ large enough.
\end{lemma}

\textit{Proof:}
Note that by the property of $(4,8)$ absorbing sets satisfied check nodes  in the substructure spanning bit nodes $(j_1,k_1)$, $(j_2,k_2)$, $(j_3,k_3)$, and $(j_4,k_4)$ can be labelled by either $(i_1,i_2,i_3,i_6,i_7,i_8)=(y,z,x,z,t,w)$ or $(i_1,i_2,i_3,i_6,i_7,i_8)=(y,x,z,w,t,z)$.

Using the technique of Section~\ref{mtrxrep} we construct the CCM for this configuration.  The binary cycle space for Fig.~\ref{figa5} has dimension $5$. We construct the following CCM by selecting the following linearly independent cycles: {$v_1-v_2-v_3,v_1-v_2-v_4,v_1-v_2-v_5,v_1-v_3-v_6-v_4,v_1-v_5-v_6-v_4$}
\begin{equation}\label{eq1b000}
\mathbf{M} =\left[
\begin{array}{cccccc}
x-t & t-z & 0 & 0 & 0\\
0 & z-w & w-y & 0 & 0 \\
x-i_5 & 0 & 0 & i_5-i_4 & 0 \\
0 & i_9-w & w-i_{10} & 0 & i_{10}-i_9 \\
0 & 0 & y-i_{10} & i_{11}-i_4& i_{10}-i_{11}
\end{array}
\right].
\end{equation}
\normalsize
The determinant of the CCM is
\small
\begin{equation}\label{eq1b111}
\begin{split}
&\det{\mathbf{M}}\\
=&[-(z-w)(i_5-i_4)((w-i_{10})(i_{10}-i_{11})-(i_9-i_{10})(i_{10}-i_9))\\
&+(i_9-w)(w-y)(i_5-i_4)(i_{10}-i_{11})](x-t)\\
&-(x-i_5)(t-z)(w-y)(i_{10}-i_9)(i_{11}-i_4) \mod p
\end{split}
\end{equation}
\normalsize

In fact this determinant always evaluates to zero for every value of $p$ for 6 non-isomorphic edge labellings. For each such labelling, once say the values of $j_1,k_1$ and $j_2$ are selected (which can be done in $p^2(p-1)$ ways), the rest of values in the configuration follows uniquely. Therefore
there are $6p^2(p-1)$ such absorbing sets.

For certain small values of $p$, $p \in \{2,3,7,11,13,31,47\}$, ~\eqref{eq1b111} has additional solutions, raising the total number of solutions to $8p^2(p-1)$. Nonetheless, the scaling $\Theta(p^3)$ of the cardinality of the absorbing sets still holds.

In the latter case we likewise establish a matrix relating the labels of the check nodes ($i_7,i_8,i_9,i_{10},i_{11}$ in Fig.~\ref{figa4}) incident to bit nodes $(j_5,k_5)$ and $(j_6,k_6)$ to adjacent checks.
Again, by imposing the bit consistency conditions, we conclude that the above constraint cannot hold for $p>23$ and therefore such a labelled configuration is in fact not possible for large enough $p$.
\hfill$\blacksquare$

As in the previous case, the variable-node of the configuration shown in Fig.~\ref{figa5} is a subgraph of the VN graph of $ (4,8) $ absorbing sets. The following again is an easy consequence of Corollary~\ref{cor:sr48}.
\begin{corollary}\label{coro1b}
If $ (4,8) $ absorbing sets are absent, the configuration shown in Fig.~\ref{figa5} is not possible in the SR SCB code.
\end{corollary}

\begin{figure}
\centering
\includegraphics[width=0.25\textwidth]{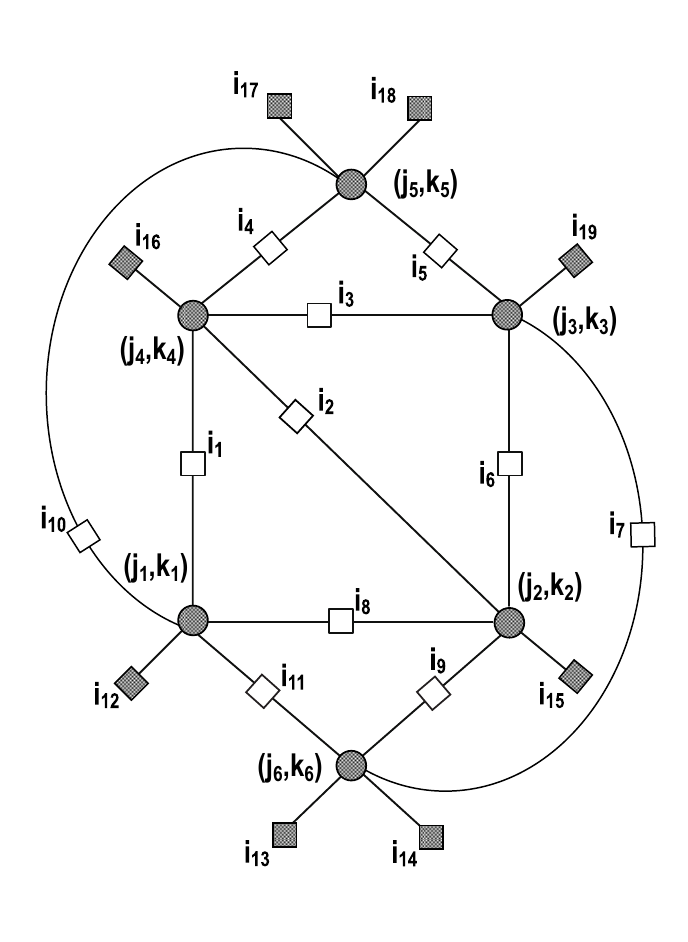}\vspace{-0.2in}
\centering
\caption{$ (6,8) $ configuration candidate 6.}\vspace{-0.2in}
\label{figa6}
\end{figure}

\subsection{$ (6,8) $ configuration candidate 6 - Fig.~\ref{figa6}}\label{app:cand6}

The last configuration we consider is the one in Fig.~\ref{figa6}.

\begin{lemma}\label{lemma1a}
In the Tanner graph corresponding to $H_{p, f(i,j)}^{5,p}$  there are no $(6,8)$ absorbing sets for $p$ large enough of the type shown in Fig.~\ref{figa6} in neither the EAB SCB nor in the SR SCB code.
\end{lemma}

\textit{Proof:}
The binary cycle space for Fig.~\ref{figa6} has dimension $6$. We construct the CCM by selecting the following linearly independent cycles: {$v_1-v_2-v_3,v_1-v_2-v_4,v_1-v_2-v_5,v_1-v_3-v_6-v_4,v_1-v_5-v_6-v_4$}
\small
\begin{equation}\label{eq1a000}
\mathbf{M}=
\left[
\begin{array}{ccccc}
i_8-i_2 & 0 & i_2-i_1 & 0 & 0\\
0 & 0 & i_1-i_4 & i_4-i_{10} & 0 \\
i_6-i_2 & i_3-i_6 & i_2-i_3 & 0 & 0 \\
0 & i_5-i_{3} & i_{3}-i_4 & i_4-i_5 & 0 \\
i_9-i_6 & i_6-i_{7} & 0 & 0 & i_7-i_9 \\
i_8-i_9 & 0 & 0 & 0 & i_9-i_{11}
\end{array}
\right].
\end{equation}
\normalsize
The inspection of $ \mathbf{M} $ reveals that $ \operatorname{rank}(\mathbf{M})=4 $ if and only if $ \det (A)=0 $ and $ \det (B) =0$, where
\begin{equation}\label{eq1a001}
A=\left[
\begin{array}{cccc}
i_8-i_2 & 0 & i_2-i_1 & 0 \\
0 & 0 & i_1-i_4 & i_4-i_{10}  \\
i_6-i_2 & i_3-i_6 & i_2-i_3 & 0 \\
0 & i_5-i_{3} & i_{3}-i_4 & i_4-i_5
\end{array}
\right],
\end{equation}
\begin{equation}\label{eq1a002}
B=\left[
\begin{array}{ccccc}
i_8-i_2 & 0 & i_2-i_1 & 0 & 0\\
0 & 0 & i_1-i_4 & i_4-i_{10} & 0 \\
i_6-i_2 & i_3-i_6 & i_2-i_3 & 0 & 0 \\
i_9-i_6 & i_6-i_{7} & 0 & 0 & i_7-i_9 \\
i_8-i_9 & 0 & 0 & 0 & i_9-i_{11}
\end{array}
\right].
\end{equation}
\normalsize
Their determinants are
\small
\begin{equation}\label{eq1a003}
\begin{array}{cc}
\det(A) &= (i_8-i_2)[-(i_1-i_4)(i_3-i_6)(i_4-i_5)\\
&+(i_4-i_{10})((i_3-i_6)(i_3-i_4)-(i_2-i_3)(i_5-i_3))]
\end{array}
\end{equation}
\begin{equation}\label{eq1a004}
\begin{array}{cc}
\det(B) &= (i_4-i_{10})[-(i_8-i_2)(i_9-i_{11})(i_2-i_3)(i_6-i_7)\\
&+(i_2-i_1)((i_6-i_2)(i_6-i_7)(i_9-i_{11})\\
&-(i_3-i_6)((i_9-i_6)(i_9-i_{11})-(i_7-i_9)(i_8-i_9)))]
\end{array}
\end{equation}
\normalsize
Then in the similar manner of the proof of lemma \ref{lemma2a}, we can show that the determinant of the corresponding matrix evaluates to zero only in finitely many choices for $p$ for either selection of $a(i)$. In particular for $a(i)=i$ it suffices for $p$ to be $>29$ and $p\neq 41$ for the configuration not to exist.
\hfill$\blacksquare$

\subsection{Proof of Lemma~\ref{lemmaqc1}}\label{pr:lemmaqc1}
For the codes with girth greater than 6, both $ (4,8) $ and $ (6,8) $ absorbing sets do not exist since these two sets \deleted[LD]{contains}\added[LD]{contain} cycle-6. For the codes with girth $ =6 $, we take $ p=31, c=5,r=6 $ as an example, with $ b=2, a=6 $. Thus the $H_{p, f(i,j)}^{5,p}$ \deleted[LD]{here} is a  sub-matrix of array code $ \tilde{H}_{p, f(i,j)}^{5,p}$ with $ f(i,j)=m(i)\cdot j$ for  $(i,m(i))$ $\in$ $\{(0,1), (1,2), (2, 4), (3,8), (4, 16)\}$. We set up the system of equations as before. For $ (4,8) $ absorbing sets, the only possible labelling for $ p=31 $ is $ (i_1, i_2, i_3, i_4, i_5, i_6) = (x,t,w,y,z,z) $ and there are five non-isomorphic solutions to the equation $ (z-w)(x-t)(y-z)-(y-w)(x-z)(z-t)\equiv 0 \mod p $: $ (x,y,z,w,t) = (2,4,1,16,8), (1,8,2,4,16), (1,2,4,16,8), (1,4,8,2,16),$ $(1,2,16,8,4) $. Each of the solutions corresponds to the matrix
\begin{equation}\label{eqqc000}
\mathbf{R}=\left[
\begin{array}{cccc}
z-x &  x-t & t-z & 0 \\
y-x  & x-z & 0 & z-y
\end{array}
\right]
\end{equation}
such that $ R(j_1,j_2,j_3,j_4)^T \equiv 0 \mod p $. Suppose the null space of each matrix is $ N_i, 1 \leq i \leq 5 $. For any  $ (4,8) $ absorbing set, $ (j_1,j_2,j_3,j_4) $ should be in $ \mathop  \cup \limits_{1 \leq i \leq 5} N_i $. Denote $ Y=\{1,6,5,30,25,26\} $, which is the index of the subcolumns in the quasi-cyclic code. $ (j_1,j_2,j_3,j_4) $ should also be in $ \mathop  \cup \limits_{i,j,k,l} (Y(i),Y(j),Y(k),Y(l)) $. However, in this case, $ \{\mathop  \cup \limits_{i,j,k,l} (Y(i),Y(j),Y(k),Y(l))\} \cap \{\mathop  \cup \limits_{1 \leq i \leq 5} N_i \} = \emptyset$. Thus $ (4,8) $ absorbing sets do not exist, and consequently $ (6,8) $ absorbing sets in Figs. \ref{cand2}, \ref{figa4} and \ref{figa5} also do not exist.

Similarly, the null space of equation~\eqref{eq2a000} and ~\eqref{eq1a000} in lemma \ref{lemma1a} and \ref{lemma2a} does not have intersection with  $ \mathop  \cup \limits_{i,j,k,l} (Y(i),Y(j),Y(k),Y(l)) $, which eliminates the possibility of existence of $ (6,8) $ absorbing sets in Fig.\ref{figa3} and \ref{figa6}.



\end{document}